\documentclass[%
 reprint,
 amsmath,amssymb,
 aps,
]{revtex4-1}

\usepackage{graphicx}
\usepackage{dcolumn}
\usepackage{bm}

\begin{document}

\title{Exotics Searches at ATLAS}%

\author{Elisa Pueschel}
\affiliation{University of Massachusetts, Amherst}
\collaboration{on behalf of the ATLAS Collaboration}


\begin{abstract}
An overview of recent searches for exotic signatures using the ATLAS detector at the LHC is given. The results presented use data collected at center-of-mass energies of $\sqrt{s}$ = 7 TeV and $\sqrt{s}$ = 8 TeV, for datasets corresponding to a variety of integrated luminosities. Resonance searches using leptons, photons, missing transverse energy, and jets are performed, as well as searches requiring custom jet and track reconstruction. No deviations from Standard Model expectations are observed.
\end{abstract}

\maketitle

\section{Introduction}
Since its formulation, the Standard Model (SM) of particle physics has enjoyed unblemished experimental success. However, the
theory leaves certain structural questions open (Why are there three generations of quarks and leptons? What is behind the observed mass hierarchy?), as well as more experimentally driven questions about the nature of dark matter. A plethora of extensions to the SM, known as ``exotic" models, have been developed in order to address these questions. This proceedings covers a variety of searches interpreted in the context of predictions from exotic models, including technicolor extensions and various Grand Unification Theories (GUTs). Supersymmetry (SUSY) is not considered, except in cases where SUSY is part of a larger extension to the SM, such as the Hidden Valley scenario. 

The measurements detailed here, using the ATLAS detector at the LHC, are searches for final state signatures that can be interpreted in the context of predictions from multiple models. Thus, the measurements are presented in terms of a search for a particular experimental signature, and then interpretations in terms of relevant models are given. This proceedings covers searches for dilepton and $WZ$ resonances; searches for seesaw heavy fermions, leptoquarks, and long-lived multi-charged particles; and several searches for lepton-jets.

\section{\label{sec:atlas}The LHC and the ATLAS Detector}

The Large Hadron Collider at CERN collided protons with center-of-mass energies of 7 TeV in 2011 and 8 TeV in 2012, corresponding to a total integrated luminosity of 4.9~$fb^{-1}$ and 20~$fb^{-1}$, respectively.

The ATLAS detector is a multi-purpose detector whose tracking, calorimetric, and muon spectrometer systems allow accurate identification and precise reconstruction of leptons, jets, photons, and missing transverse energy~\cite{detector}.

\section{\label{sec:dilepton} Dilepton Resonance Search}
A number of SM extensions predict a heavy resonance that decays to $\mu^{+}\mu^{-}$ or $e^{+}e^{-}$. For instance, GUT E$_{6}$
models and others predict a $Z'$, a heavy neutral gauge boson of spin 1~\cite{SSM, E6}. In models that use Randrall-Sundrum
extra dimensions, a spin 2 Randall-Sundrum graviton, $G^{*}$, is predicted~\cite{RSgrav}. 

Searches for resonances in the $\mu^{+}\mu^{-}$ and $e^{+}e^{-}$ invariant mass spectra are performed on the full $\sqrt{s}$ = 8
TeV dataset~\cite{dileptons}. Events in the di-muon channel are triggered with a single high $p_{T}$ muon trigger. A di-photon
trigger is used for the di-electron channel, which has comparable efficiency for signal to electron triggers, and facilitates background
determination.

After applying the trigger, a pair of high quality, opposite charge electrons or muons are selected. The main background is Drell-Yan ($Z/\gamma^{*} \rightarrow l^{+}l^{-}$), but the di-electron channel also suffers from QCD multijet and $W$+jet backgrounds, where a jet is misidentified as an electron. Both channels also have reducible backgrounds from $t\bar{t}$ and diboson processes. The expected number of events for most background sources is evaluated using simulated events, normalized to the SM predicted cross section for each process. Simulation is not used for the multijet and $W$+jet backgrounds, instead the probability for a jet to fake a lepton is measured on data control samples.

\begin{figure*}[htbp]
\includegraphics[width=0.32\textwidth]{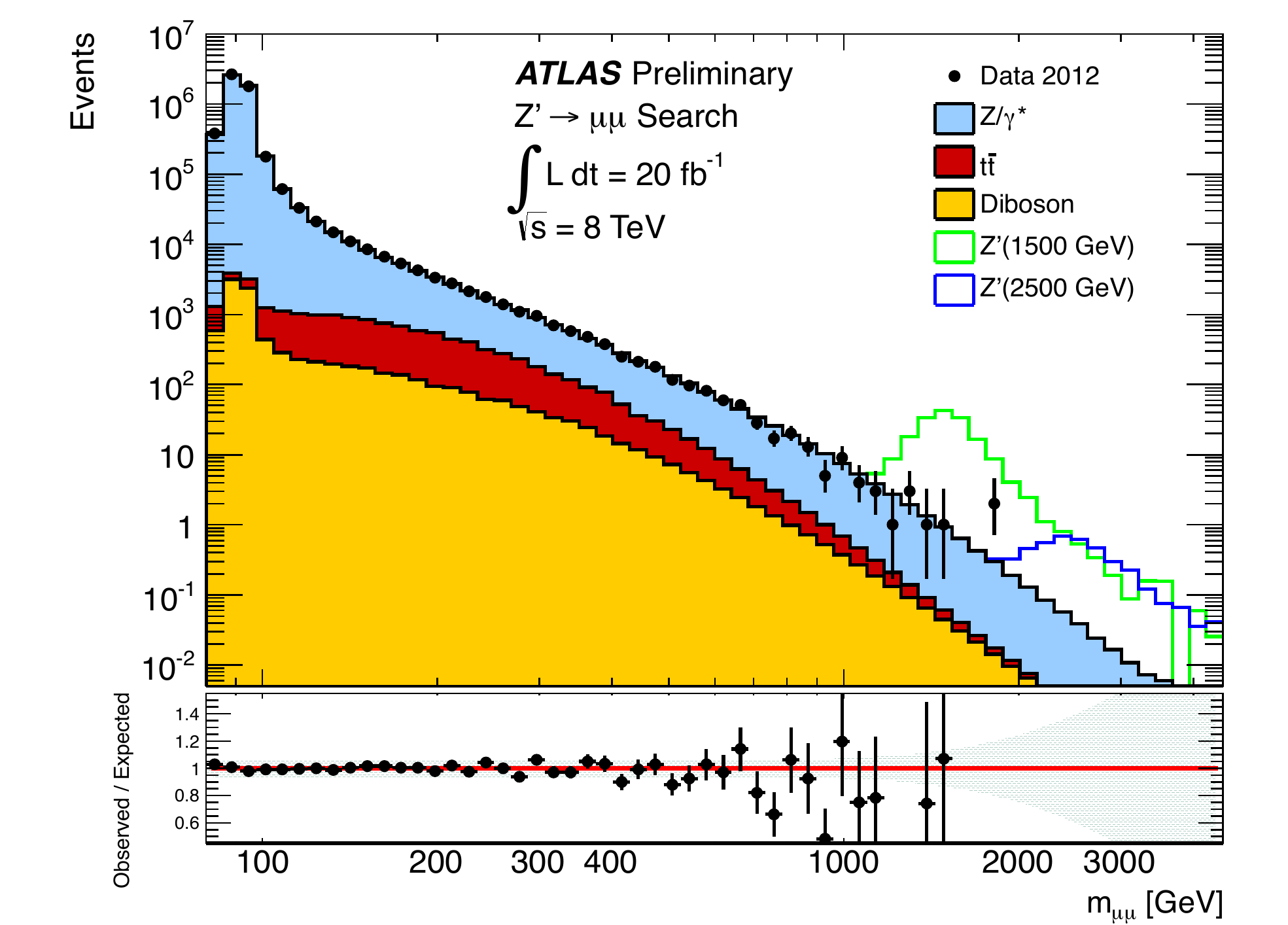}
\includegraphics[width=0.32\textwidth]{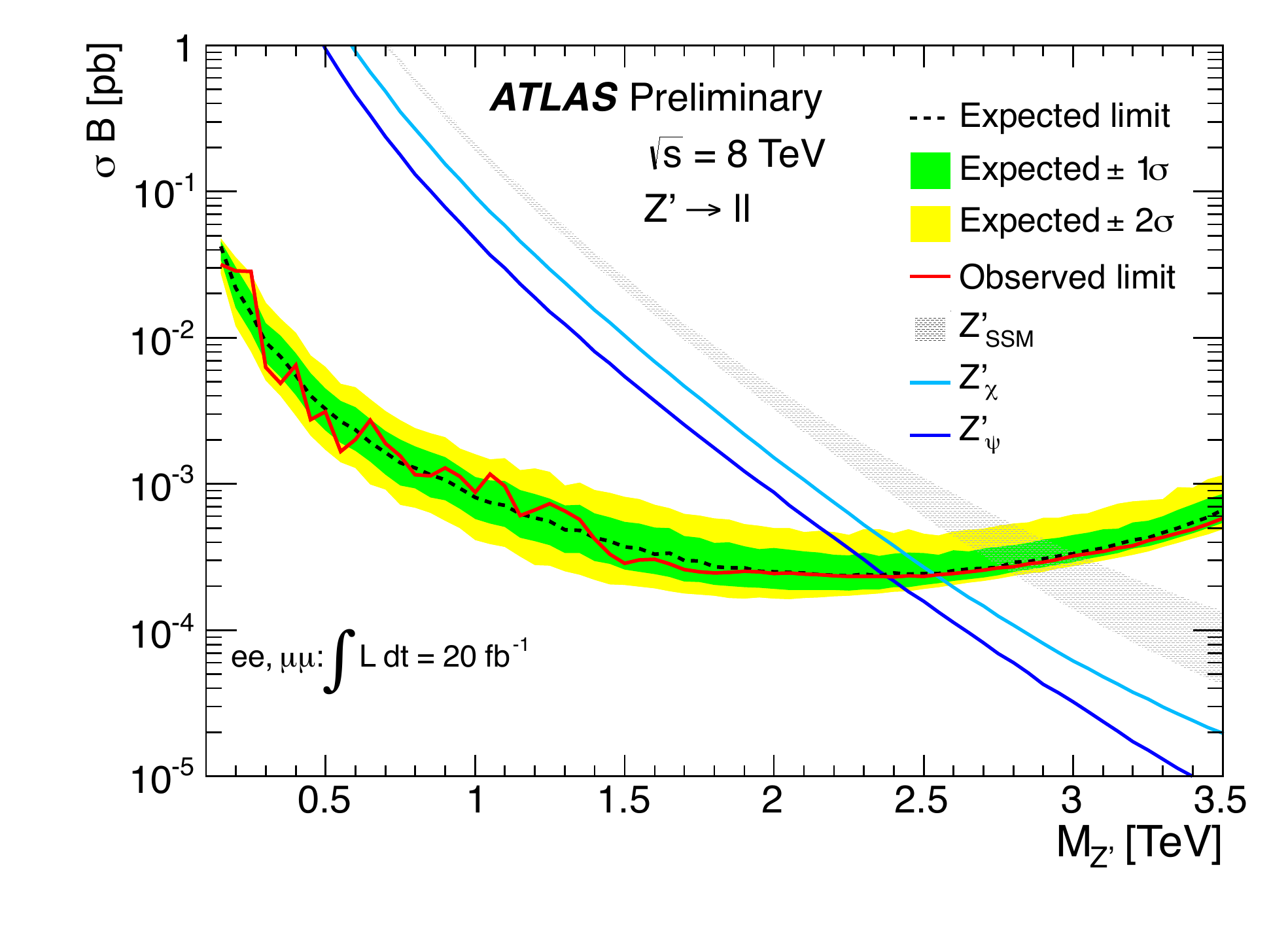}
\includegraphics[width=0.32\textwidth]{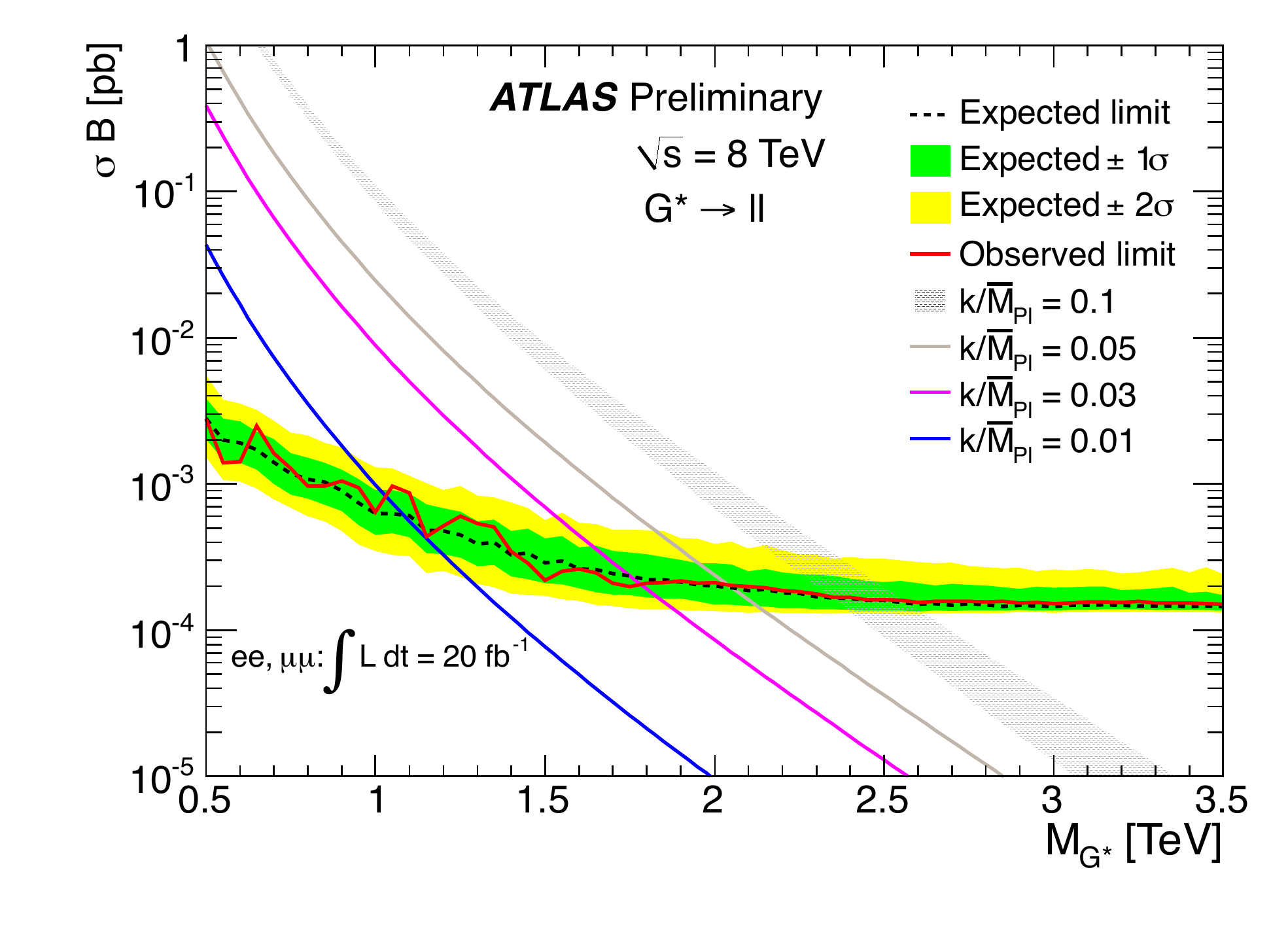}
\caption{Left: Comparison between data and expected background in the di-muon channel. Middle: Limits on $Z'$ cross section times branching ratio as a function of $Z'$ mass. Right: Limits on Randall-Sundrum graviton cross section times branching ratio as a function of graviton mass~\cite{dileptons}.}
\label{dilepton}
\end{figure*}

The total background contribution is normalized at the $Z$ mass peak in data, which removes many systematic uncertainties, though mass-dependent uncertainties remain. The left panel in Figure~\ref{dilepton} shows the comparison between data and expected background in the di-muon channel, with signal contributions from a $Z'$ with a mass of 1.5 and 2.5 TeV overlaid for illustration purposes. No excess above the background expectation is observed. Limits are then set in the context of several theoretical models. The derived limits are shown in Figure~\ref{dilepton} for several $Z'$ models (middle) and for the Randall-Sundrum graviton for different k/$\bar{M}_{Pl}$ couplings (right). The lower limits are set at 2.86 TeV for a sequential Standard Model (SSM) $Z'$, 2.38-2.54 TeV for E$_{6}$ models, and 2.47 TeV for a Randall-Sundrum graviton at 95\% confidence level (CL). These represent significant improvements in the exclusion limits compared to previous results.

\section{\label{sec:WZres} $WZ$ Resonance Search}
Another standard ``bump-hunt" is the search for a $WZ$ resonance. A number of SM extensions, including extended gauge models, composite Higgs/Little Higgs models, and low scale technicolor, predict a resonant $W'$ or technirho $\rho_{T}$. In this search, a $WZ$ resonance from a $W'$ or $\rho_{T}$ decay is probed in the channel $WZ \rightarrow l\nu l'l'$. An 8 TeV dataset corresponding to 13~$fb^{-1}$ of integrated luminosity is used~\cite{WZres}.

Events are triggered based on the presence of high $p_{T}$ muons or electrons. The final states may consist of $\mu \nu \mu^{+} \mu^{-}$, $\mu \nu e^{+} e^{-}$, $e \nu \mu^{+} \mu^{-}$, or $e \nu e^{+} e^{-}$. Therefore, events with three leptons of $p_{T}>$25 GeV and missing transverse energy ($E^{miss}_{T}$) $>$25 GeV are selected, and events with a fourth lepton of $p_{T}>$20 GeV are rejected. Two of the leptons are required to have the same flavor and opposite charges, and the pair must have an invariant mass consistent with the $Z$ mass. The remaining lepton is combined with $E^{miss}_{T}$ to form the $W$ transverse mass, $m^{W}_{T} = \sqrt{2 p^{l}_{T} E^{miss}_{T}(1 - cos\Delta \phi)}$, where $\Delta \phi$ is the opening angle between the lepton and the $E^{miss}_{T}$. Events are selected with $m^{W}_{T} <$ 100 GeV.

There are several sources of background in this final state. The SM $WZ$ production is an irreducible background. Reducible background contributions come from $ZZ$ decays to four leptons, where one lepton fails the selection requirements, $Z\gamma$ production, where the photon is misidentified as an electron, and $ll$+jets production, where either a jet is misidentified as a lepton, or a hadron decays to produce a lepton in the final state. The diboson backgrounds are estimated using simulated events, although the large $WZ$ background is validated against a $WZ$ data control sample. The probability for $ll$+jets events (which include $Z$+jets, $t\bar{t}$ and $Wt$) to fake the signal is estimated from data, and the contribution is cross-checked against a $Z$+jets data control sample.
 
\begin{figure*}[htbp]
\includegraphics[width=0.32\textwidth]{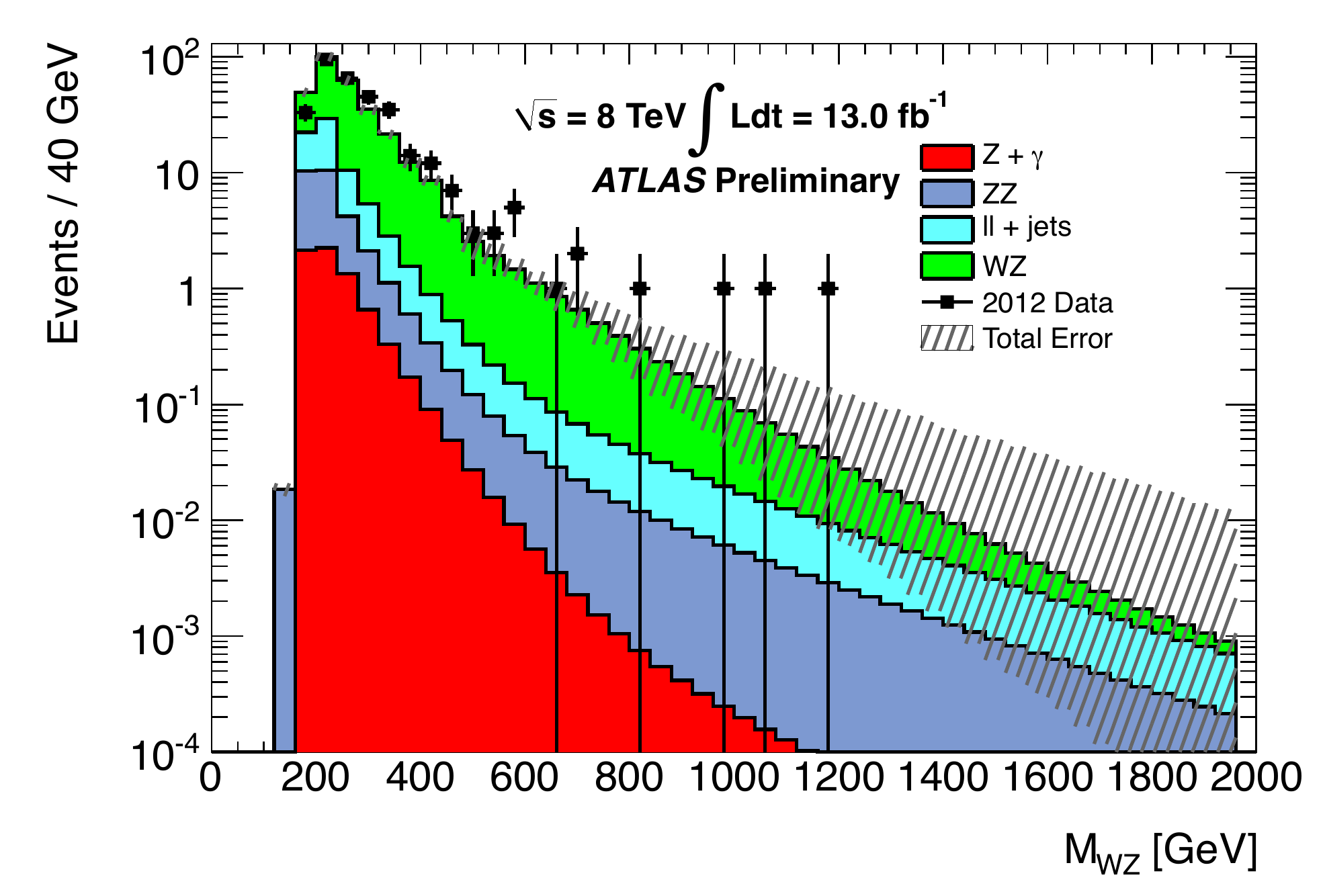}
\includegraphics[width=0.32\textwidth]{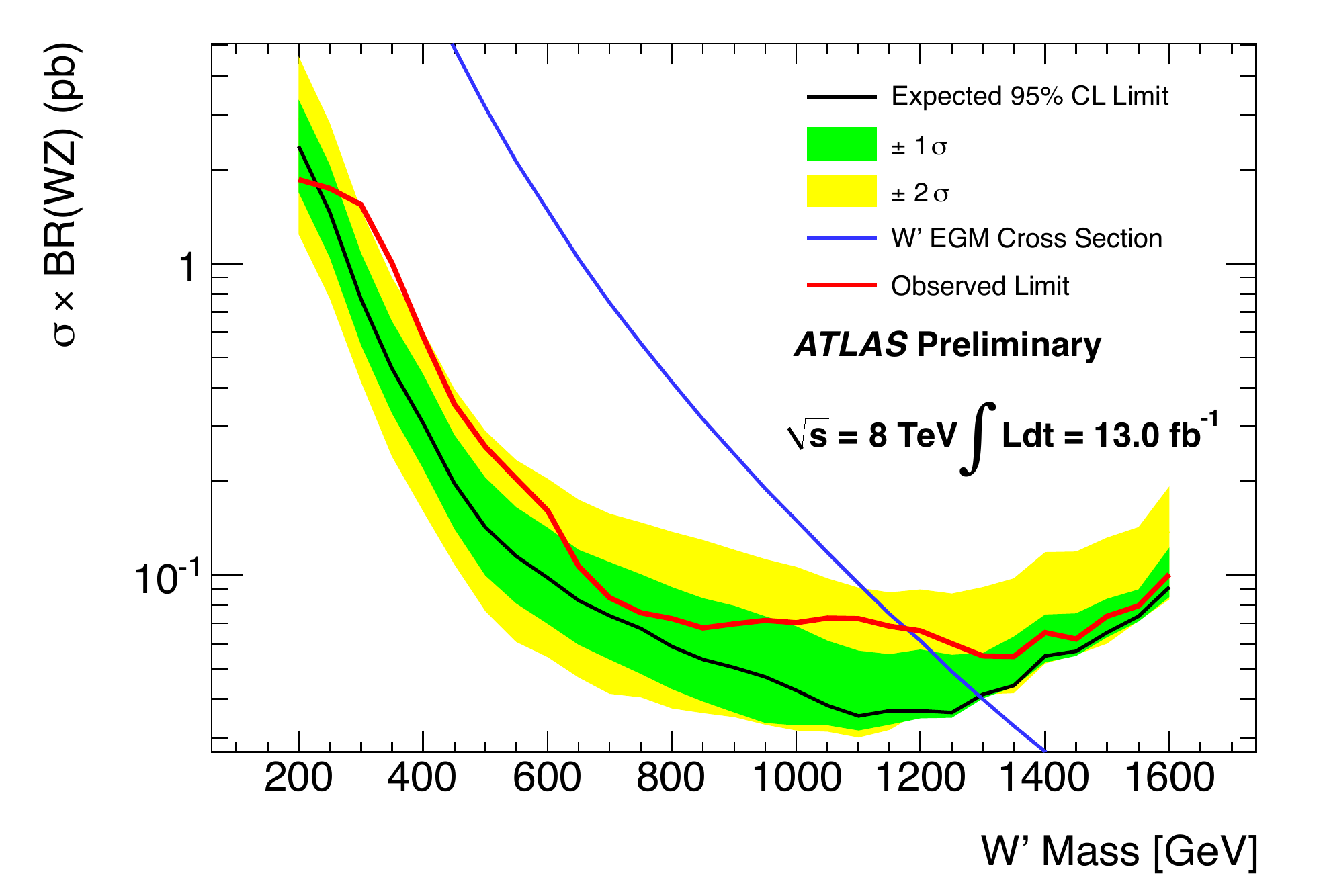}
\includegraphics[width=0.32\textwidth]{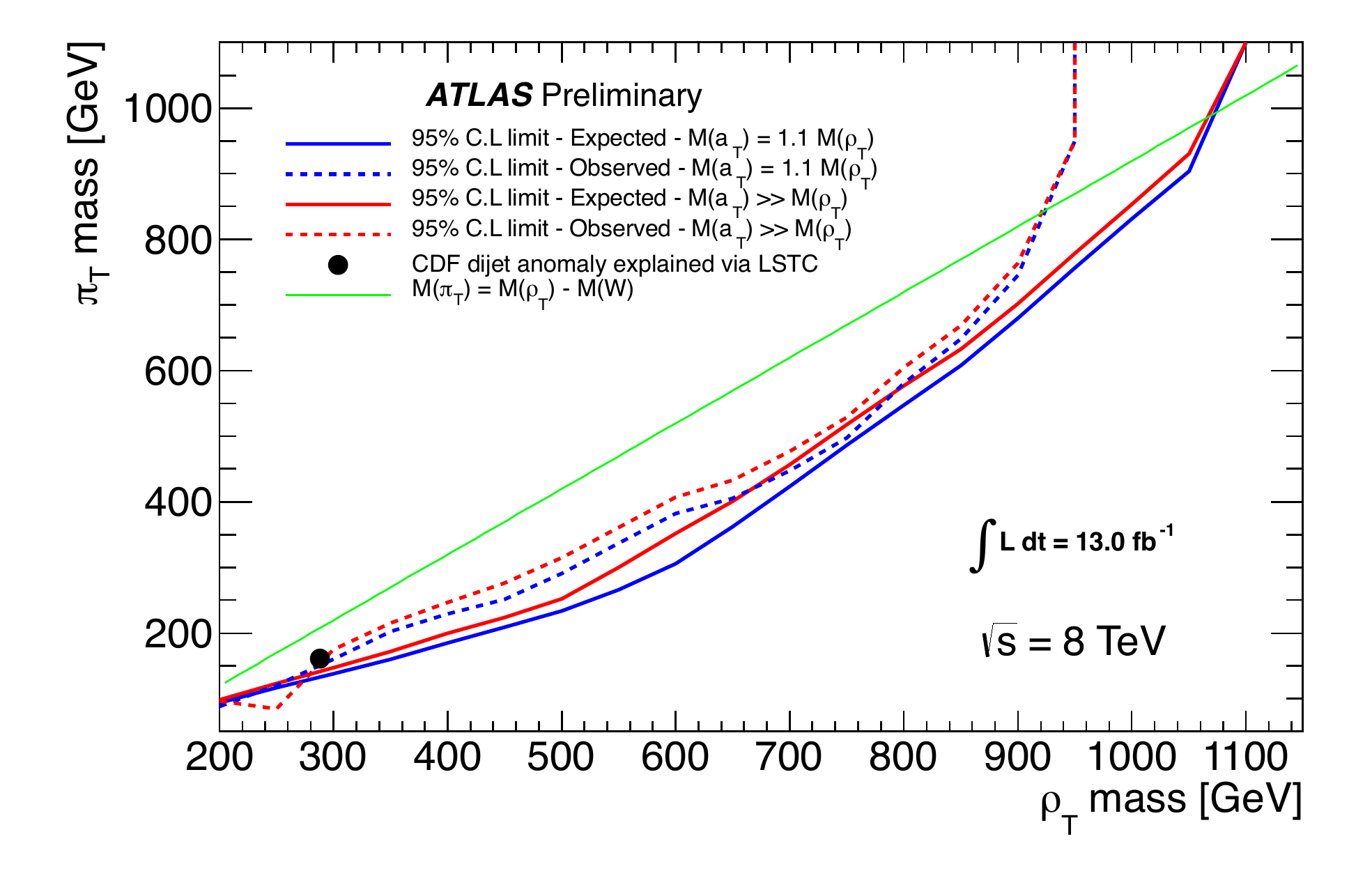}
\caption{Left: Comparison between data and expected background for $WZ$ invariant mass. Middle: Limits on $W'$ cross section times branching ratio to $WZ$ as a function of $W'$ mass. Right: Exclusions region in the $\rho_{T} - \pi_{T}$ plane~\cite{WZres}.}
\label{Wprime}
\end{figure*}

 In Fig.~\ref{Wprime}, left panel, data is compared with the predicted background for the $WZ$ invariant mass. A fit is performed to extract the number of expected background events in the tail of the distribution, using a single exponential for $WZ$ background, and a double exponential for non-$WZ$ backgrounds. No discrepancy is seen between the observed number of events and the predicted background, and limits combining all four lepton final states are set on the $W'$ cross section times branching fraction to $WZ$. The allowed and excluded regions in the technirho/technipion mass plane are also determined. The limits and exclusion region are shown in Fig.~\ref{Wprime} for the $W'$ (middle panel) and $\rho_{T}$ (right panel). An extended gauge model $W'$ is excluded up to 1180 GeV at 95\% CL, and a $\rho_{T}$ is excluded up to 920 GeV.

\section{\label{sec:seesaw} Search for Type III Seesaw Heavy Fermions}
Seesaw models provide a mechanism for giving mass to neutrinos, as well as forcing their masses to be small, concordant with experimental measurements. Majorana terms are added to the SM Lagrangian, and the new particles have large masses that suppress the SM neutrino masses. There are three types of seesaw mechanism, of which Type II and Type III predict new particles in an energy range accessible at the LHC~\cite{type2seesaw, type3seesaw}. In the Type III mechanism, at least two new fermionic triplets are added to the SM, generating the neutrino masses. The lightest triplet consists of a charged pair, $N^{\pm}$, and a $N^{0}$ neutral heavy fermion. All three particles have the same mass and couple to gauge bosons. A search is performed for the production of $pp\rightarrow N^{\pm}N^{0}$, with the heavy fermions decaying to $Wl$ and $Zl$, respectively.

The search was performed on an 8 TeV dataset corresponding to an integrated luminosity of 5.8~$fb^{-1}$~\cite{seesaw}. The final state in this case corresponds to a multilepton signature, and events are triggered using single muon and single electron triggers.  Four leptons, with $p_{T}>$ 25 GeV for the leading lepton and $p_{T}>$ 10 GeV for the three subleading leptons, are required, which helps suppress $Z$+jets and $WZ$ backgrounds. One pair of leptons is required to be consistent with a $Z\rightarrow ll$ decay, with opposite charges, same flavor, and an invariant mass within 20 GeV of the $Z$ mass. Events with a second $Z$ candidate are vetoed in order to reject $ZZ$ background. The $Z$ candidate and third lepton are then used to reconstruct the invariant mass of $N^{\pm}$, $m_{Z(ll)l'}$.

The dominant SM background is $ZZ$, with further contributions from $ZZZ^{*}$, $ZWW^{*}$, $t\bar{t}V$, and $Z$+jets. Data control regions are used to compare data and simulation for the two largest background processes, $ZZ$ and $Z$+jets. For the $ZZ$ control region, the veto on the second $Z$ was reversed, and for the $Z$+jet control region, isolation requirements were loosened and impact parameter requirements were removed. After finding good agreement between data and simulation for $m_{Z(ll)l'}$ in the control regions, the normalization for the background was taken from simulation in the signal region.

\begin{figure*}[htbp]
\includegraphics[width=0.49\textwidth]{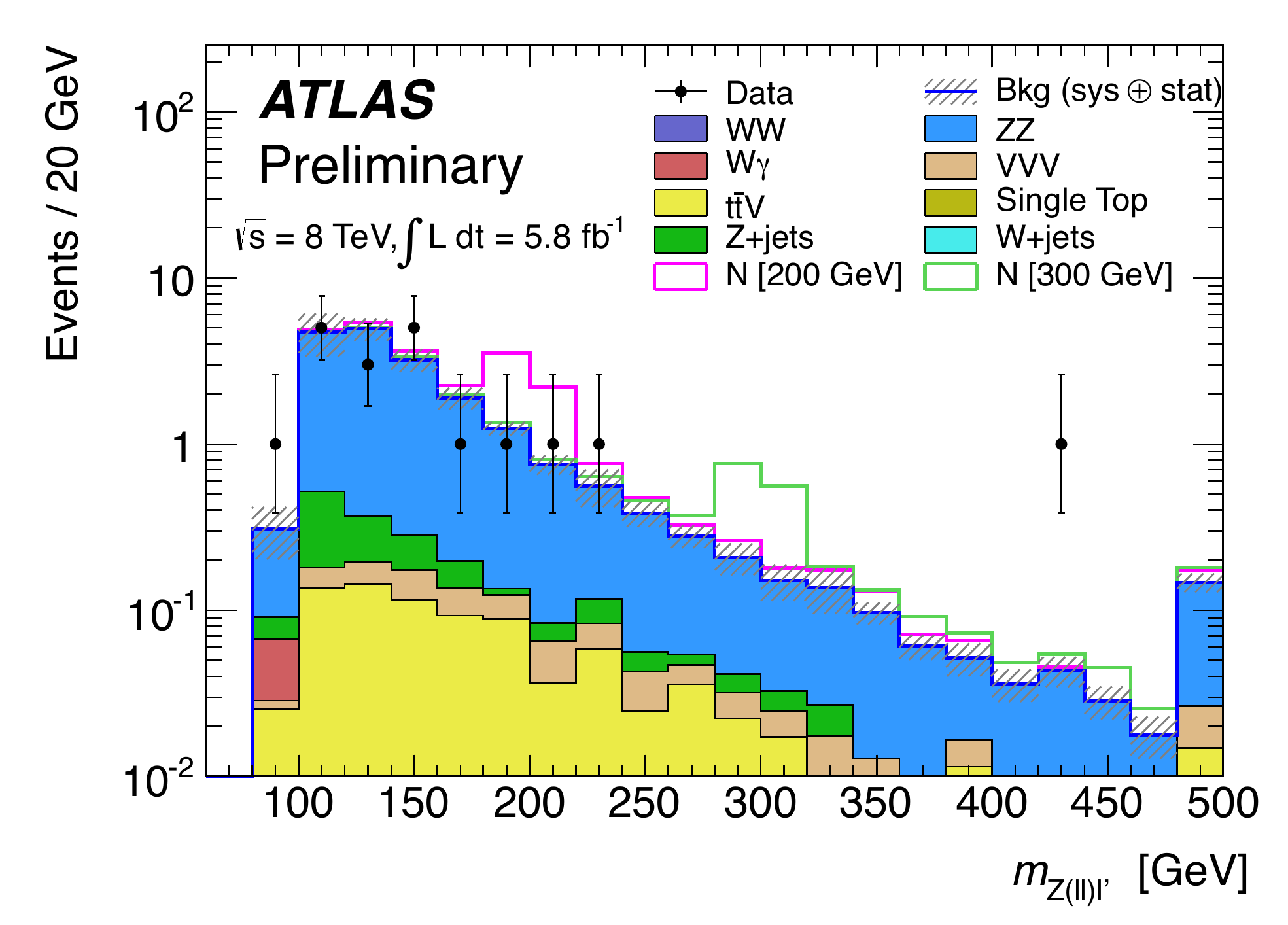}
\includegraphics[width=0.49\textwidth]{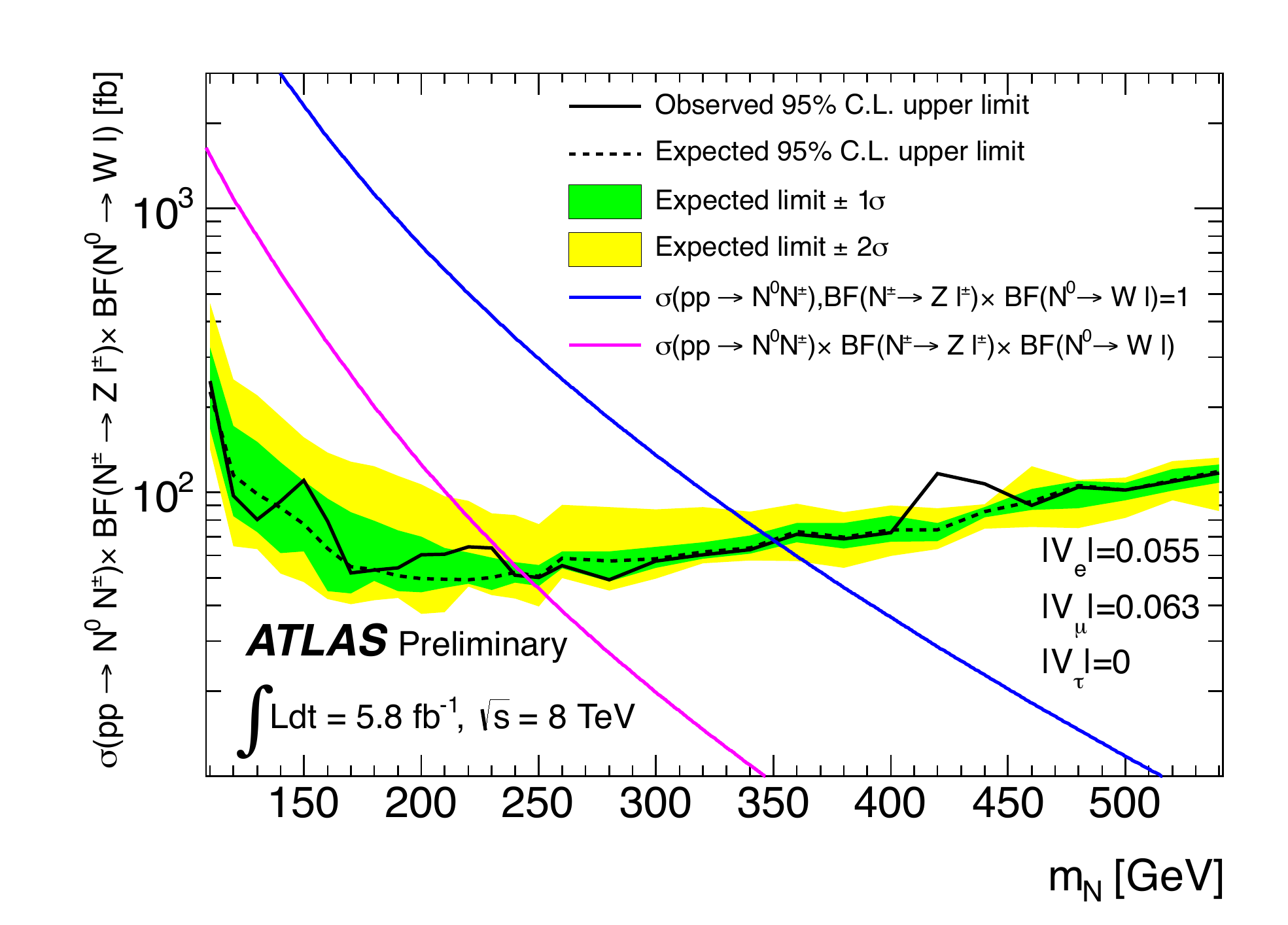}
\caption{Left: Comparison between data and expected background for $m_{Z(ll)l'}$ invariant mass. Right: Limit on the cross section times branching fraction, $\sigma(pp)\rightarrow N^{0}N^{\pm} \times BF(N^{\pm}\rightarrow Zl) \times BF(N^{0}\rightarrow Wl)$ as a function of the mass of the Type III seesaw fermion~\cite{seesaw}.}
\label{seesaw}
\end{figure*}

The agreement between data and background simulation for $m_{Z(ll)l'}$ is shown in Fig.~\ref{seesaw} (left panel), with signal shapes for the resonant $N^{\pm}$ overlaid for illustration. The data do not deviate significantly from the background prediction. A limit on the cross section times branching fraction, $\sigma(pp)\rightarrow N^{0}N^{\pm} \times BF(N^{\pm}\rightarrow Zl) \times BF(N^{0}\rightarrow Wl)$ is set as a function of the heavy fermion mass, as shown in Fig.~\ref{seesaw} (right panel). The limit is calculated for values of  $BF(N^{\pm}\rightarrow Zl) \times BF(N^{0}\rightarrow Wl)$ between 0.15 and 1.00. For the most stringent case, a 95\% CL lower limit on the mass of $N$ is set at 245 GeV.

\section{\label{sec:leptoquark} Leptoquark Search}
The final resonance-type search discussed here is a search for third generation leptoquarks decaying as $LQ3 \rightarrow \tau b$. Leptoquarks carry both lepton and baryon numbers and can unify the quark and lepton sectors. Such particles are predicted by an number of SM extensions, including technicolor models and GUTs. They are hypothesized to consist of three generations, in which each generation of leptoquarks only couples to a single generation of quarks and leptons. Thus, the third generation leptoquark $LQ3$ decays with a 100\% branching fraction to $\tau b$. Leptoquarks would be pair produced in $pp$ collisions, like quarks, resulting in a $\tau b \tau b$ final state.

The search was performed on a 7 TeV dataset corresponding to 4.7~$fb^{-1}$ of integrated luminosity~\cite{leptoquark}. Each of the two taus in the final state can decay leptonically or hadronically, leading to four possible final states. The leptonic-hadronic combination is used for this measurement, as it has the highest branching fraction and single lepton triggers can be used for event selection. Thus, the final state consists of a muon or electron, two $b$ jets, the visible decay products of the hadronically decaying tau, and missing transverse energy from three neutrinos. 

Events are triggered with a single muon or electron trigger, and events are required to have one and only one electron or muon with $p_{T} > $25 GeV or 20 GeV, respectively. An opposite sign hadronically decaying tau of $p_{T} >$ 30 GeV is also required. Two jets are required, with $p_{T} >$ 50 GeV for the leading and $p_{T} >$ 25 GeV for the sub-leading, and at least one is required to be tagged as a $b$ jet. Finally, more than 20 GeV of missing transverse energy is required, and requirements are imposed on the opening angle between the $E^{miss}_{T}$ and visible decay products of the hadronically decaying tau.

The background in this analysis is dominated by $t\bar{t}$ decays, and receives contributions from $W$+jets, $Z/\gamma^{*}$+jets, single top, diboson and multijet production processes. The normalizations of the background processes are derived from data in control regions.
 
\begin{figure*}[htbp]
\includegraphics[width=0.32\textwidth]{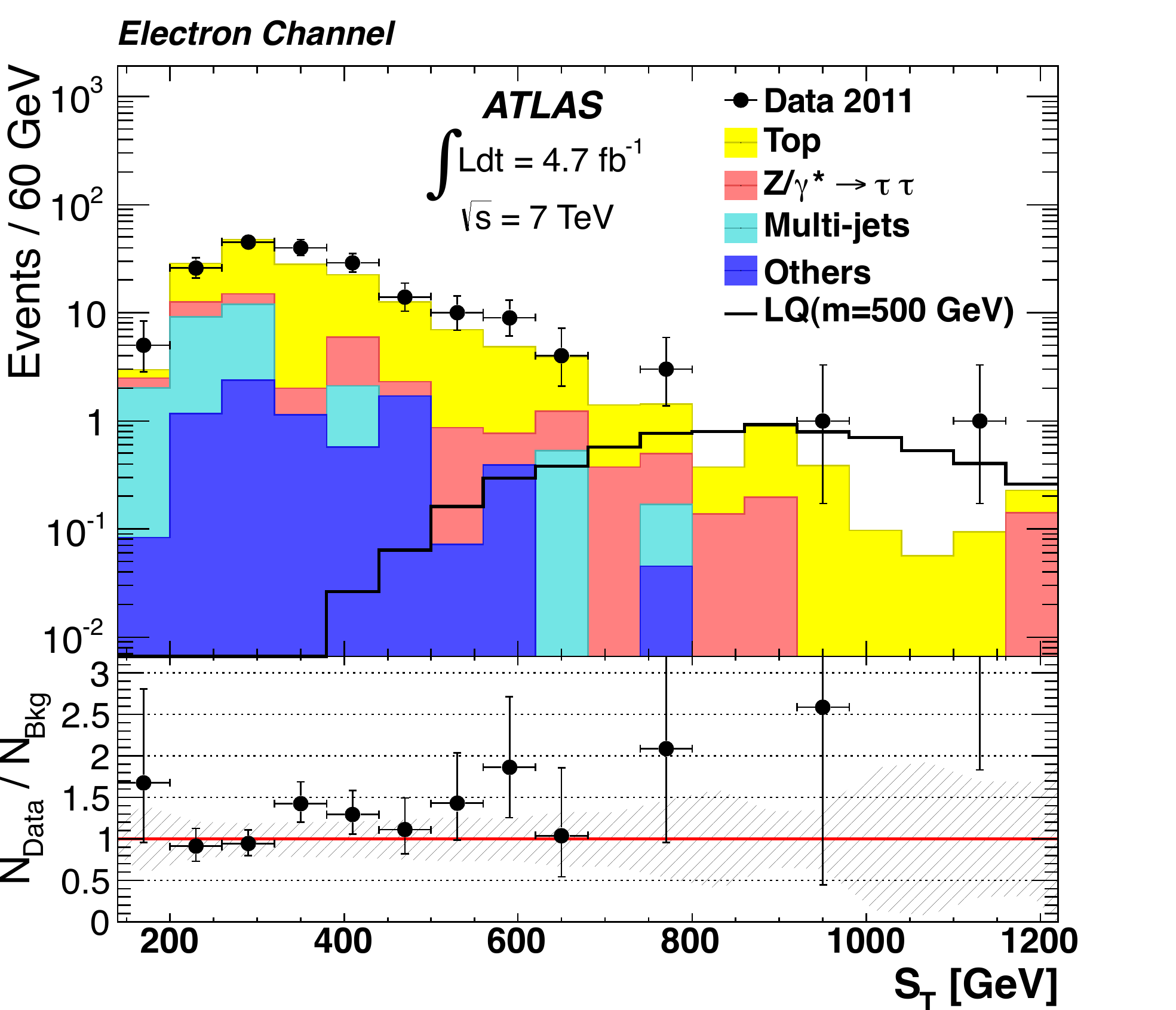}
\includegraphics[width=0.32\textwidth]{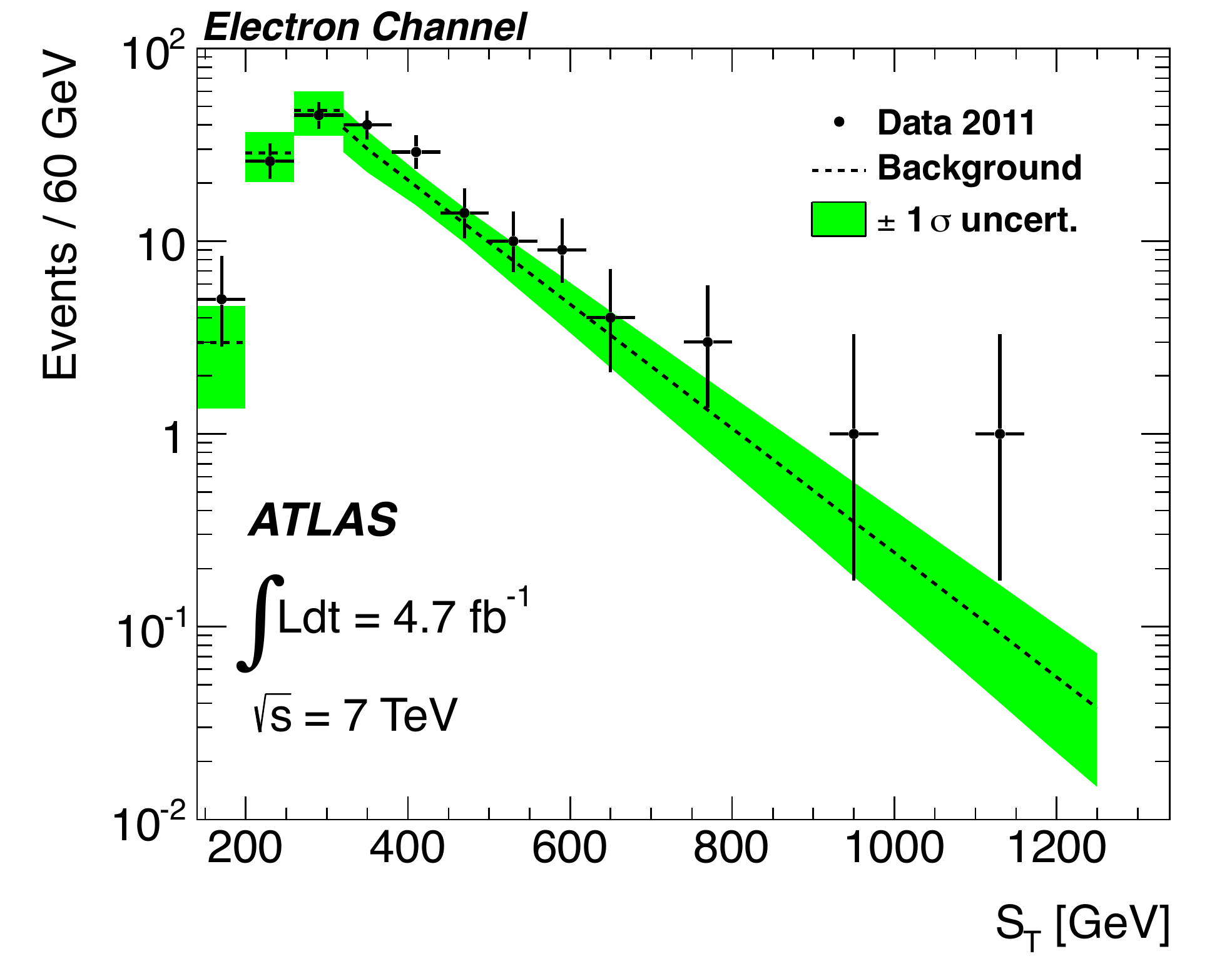}
\includegraphics[width=0.32\textwidth]{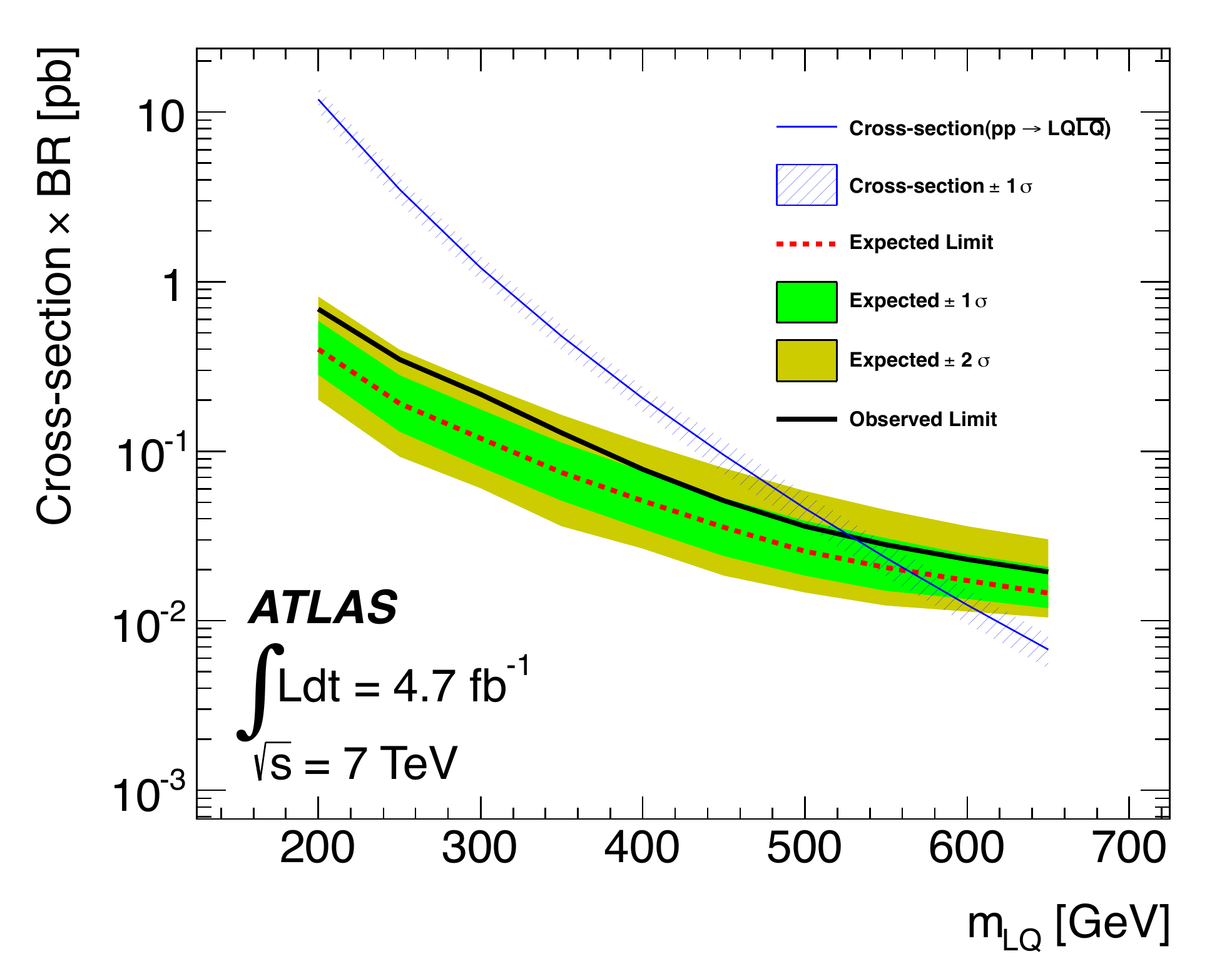}
\caption{Left: Comparison between data and background prediction for the variable $S_{T}$ in the electron channel. Middle: Fit of the background prediction. Right: Combined limit from muon and electron channels on cross section times branching ratio for a third generation leptoquark, as a function of leptoquark mass~\cite{leptoquark}.}
\label{leptoquark}
\end{figure*}

 Data and background predictions are compared for the variable $S_{T}$, which is a scalar sum of the momenta of the physics objects reconstructed in the event: $S_{T} = p^{e/\mu}_{T} + p^{\tau}_{T} + p^{jet1}_{T} + p^{jet2}_{T} + E^{miss}_{T}$. The comparison is shown in the left panel of Fig.~\ref{leptoquark} for the electron channel. The simulated background distribution for $S_{T}$ is fit in order to provide a smooth background expectation up to the high $S_{T}$ tail, where there are few events (Fig.~\ref{leptoquark} middle panel). The data are consistent with a background-only hypothesis, and the results are translated into 95\% CL limits on $LQ3$'s production cross section times branching ratio as a function of $m_{LQ3}$. Limits are set individually in the electron and muon channels, and then combined into a single limit, shown in the right panel of Fig.~\ref{leptoquark}. Third generation leptoquarks with masses below 534 GeV are excluded.
 
\section{\label{sec:llmch} Search for Long-lived Multi-charged Particles}
The search for long-lived multi-charged particles is a generic search for long-lived objects with exotic ionization signatures. Such an object could be a dyon, a micro black hole, or a $Q$-ball, objects predicted in a number of extensions to the SM. This measurement considers particles that pass through the entire ATLAS detector, interacting from the inner detector to the muon spectrometer. Particles with electric charge in the range $|q|$ = 2e - 6e are considered, which would produce distinctive ionization signatures, as specific energy loss depends quadratically on particle charge. The particles are assumed to be produced via the Drell-Yan process. A 7 TeV dataset corresponding to an integrated luminosity of 4.4~$fb^{-1}$ is used~\cite{llmch}.

\begin{figure*}[htbp]
\includegraphics[width=0.32\textwidth]{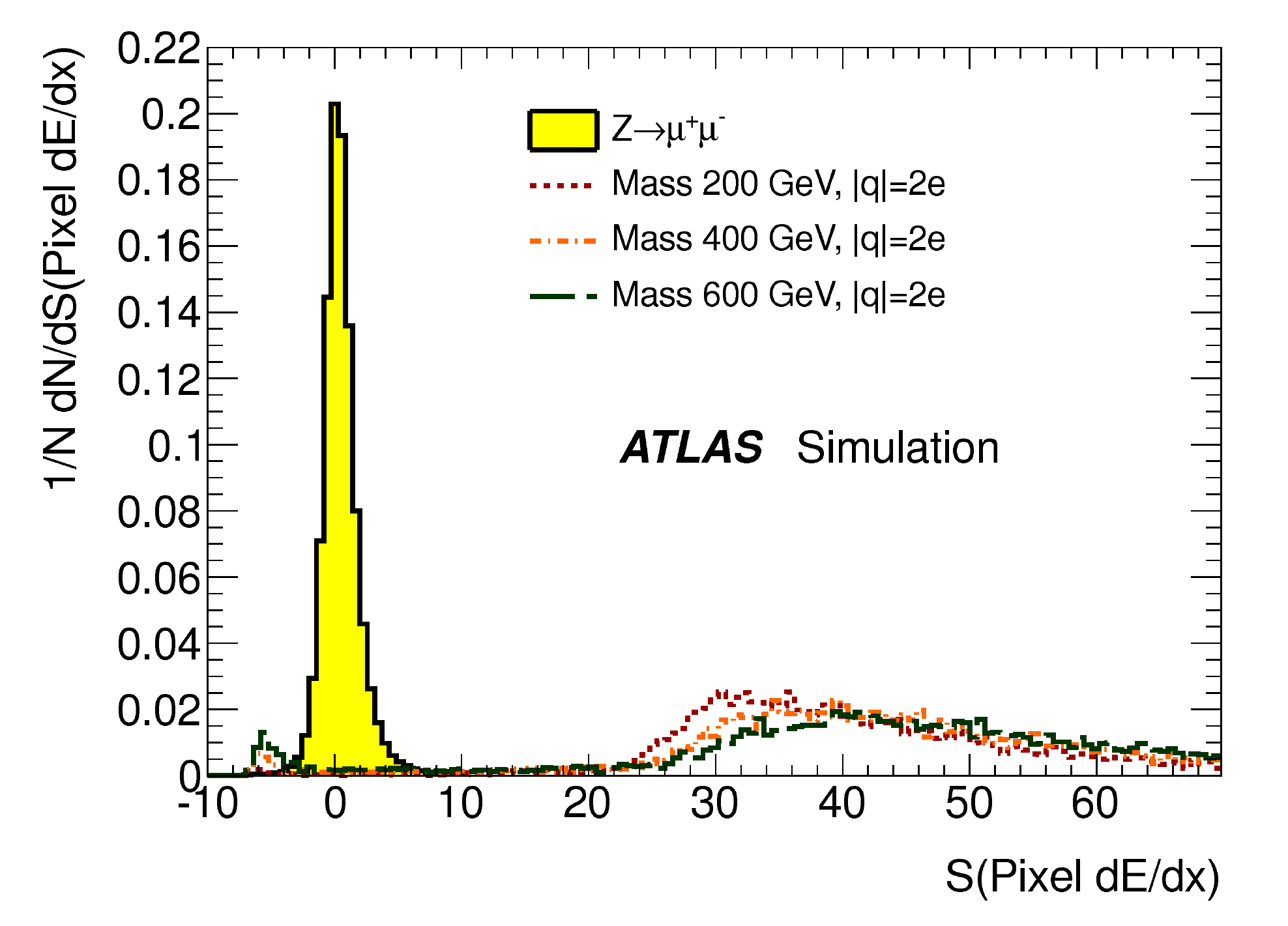}
\includegraphics[width=0.32\textwidth]{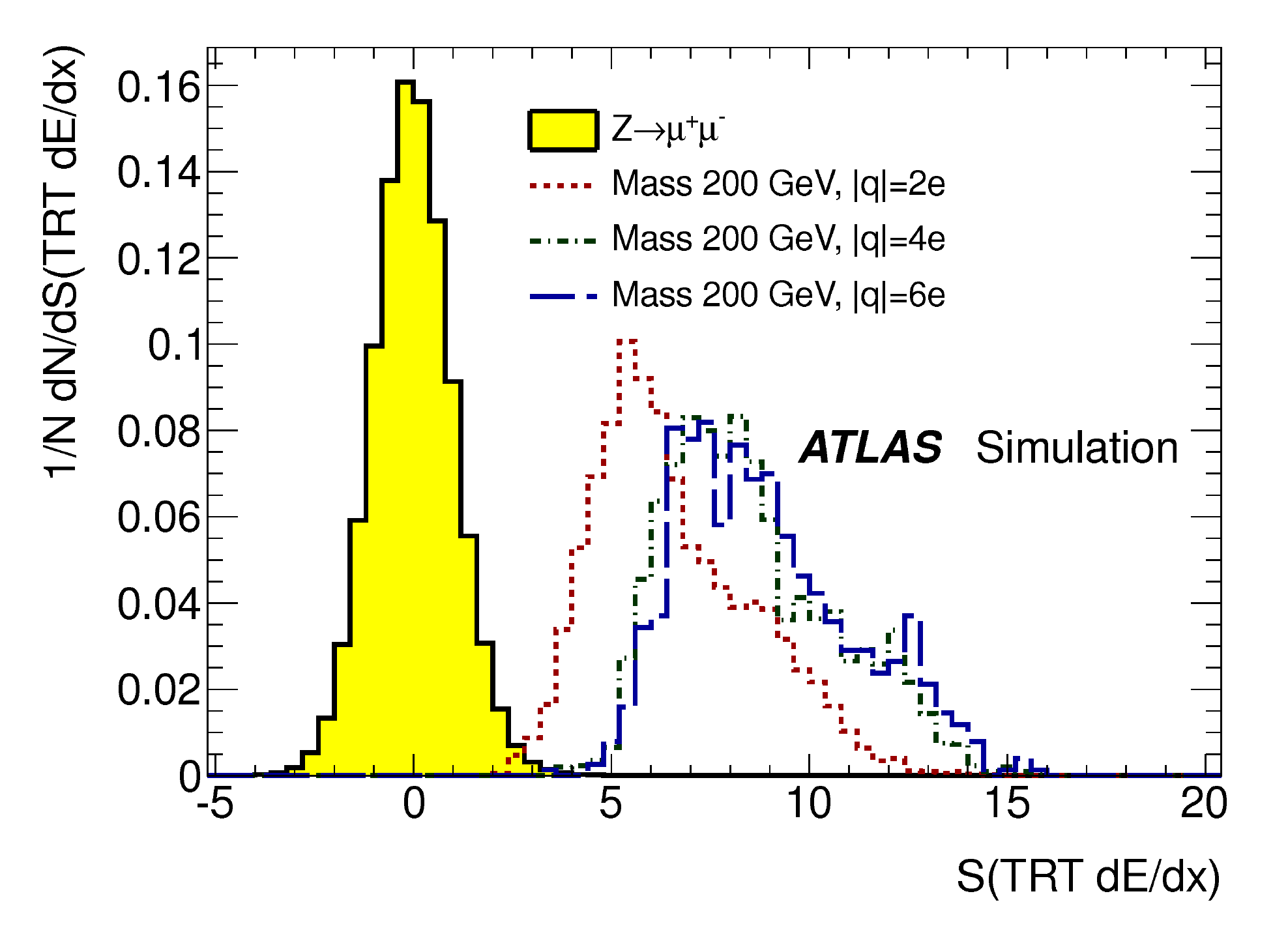}
\includegraphics[width=0.32\textwidth]{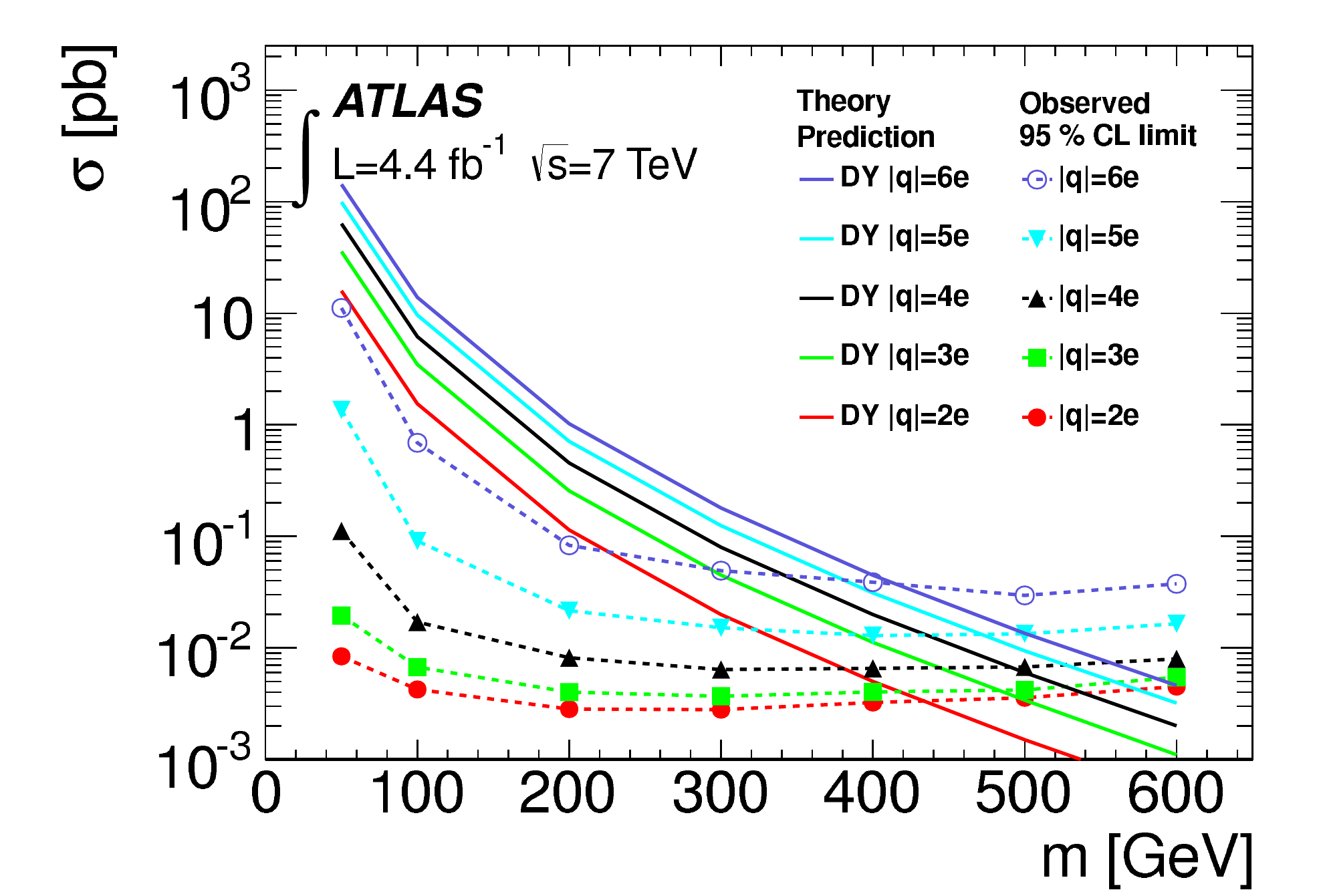}
\caption{Left, Middle: Pixel and TRT $dE/dx$ significance for simulated signal and muons from $Z\rightarrow \mu^{+}\mu^{-}$ decays in data. Right: Limits on the long-lived, multi-charged particle cross section as a function of mass for different particle charges, assuming Drell-Yan production~\cite{llmch}.}
\label{llmch}
\end{figure*}

In order to quantify the signal's anomalous ionization for the purpose of selection, the ionization for simulated signal is compared with that of muons from $Z\rightarrow \mu\mu$ decays. The difference in specific energy loss, $dE/dx$ is used to form a significance estimator:

\begin{equation}
S(dE/dx) = \frac{dE/dx_{signal} - <dE/dx_{mu}>}{\sigma(dE/dx_{\mu})}
\end{equation}
where $<dE/dx_{mu}>$ is the mean muon $dE/dx$ and $\sigma(dE/dx_{\mu})$ is the quantity's resolution. The $dE/dx$ significance is estimated in two parts of the inner detector, the pixel region and the transition radiation tracker (TRT), as well as in the muon spectrometer, in the muon drift tubes (MDT). The distributions of $S(dE/dx)$ for pixel and TRT are shown in Fig.~\ref{llmch} in the left and central panels, and the separation between signal and muons is clear. Separation is good for the MDTs, as well, and further separation can be gained based on the fraction of high threshold hits in the TRT, $f^{HT}$.

Event selection starts with the single muon trigger, which requires a muon with $p_{T} >$ 18 GeV. Candidate long-lived multi-charged particles are selected from the collection of particles initially identified as muons, either as a individual muon of $p_{T} >$ 75 GeV or as a pair of $p_{T} >$ 15 GeV muons. The $dE/dx$ significance of these candidates is then analyzed. Candidates are pre-selected as $|q|$ = 2e or $|q| >$ 2e based on $f^{HT}$ and the pixel $dE/dx$. Then, further selection on the TRT and MDT significance is applied: $S$(MDT $dE/dx$) $>$ 3 and $S$(TRT $dE/dx$) $>$ 4 for $|q|$ = 2e and $S$(MDT $dE/dx$) $>$ 4 and $S$(TRT $dE/dx$) $>$ 5 for $|q| >$ 2e. 

The background is then estimated using the ABCD method, in which the selections on the MDT and TRT significance are reversed, both individually and together, to produce three control samples. The amount of background in the resulting control regions is then extrapolated into the signal region.

The predicted background in both the $|q|$ = 2e and $|q| >$ 2e signal regions was consistent with the observed number of events. Consequently, limits on the multi-charged particle cross section as a function of mass were set for each value of $|q|$ between 2e and 6e, as shown in the right panel of Fig.~\ref{llmch}. The excluded mass range depends on the signal charge, starting at 50 GeV and extending to 430, 480, 490, 470 and 420 GeV for $|q|$ = 2e, 3e, 4e, 5e and 6e, respectively.

\section{\label{sec:leptonjet} Lepton-Jet Searches}
Recent measurements of an astrophysical positron excess by cosmic-ray and dark matter direct detection experiments have excited model building to explain the excess. Popular models have focused on the possibility of a hidden sector of particles that only couple to SM particles via a massive mediator. Such hidden sectors are natural consequences of some SUSY and string theory models. Should the mediator to the hidden sector be produced at the LHC, it could decay to hidden sector particles, including to a ``dark photon" that mixes with the SM photon. The dark photon is predicted to decay to electron-positron pairs, which are heavily boosted and thus manifest a collimated, jet-like structure. Due to the absence of a proton excess in the cosmic ray spectrum, the dark photon is expected to have a mass less than $\sim$2~GeV, below the threshold for proton-antiproton production.

\subsection{\label{sec:lj1}Prompt electron-jets from a Higgs decay}
In this measurement, the Higgs acts as mediator to the hidden sector, decaying to hidden sector particles $h_{d,1}$ and $h_{d,2}$, either directly, or in a cascade. The dark sector particle $h_{d,2}$ can then decay to electron-jets. Simulated samples were generated at Higgs masses of 100, 125, and 140~GeV, and dark photon $\gamma_{d}$ masses of 100 and 200~MeV.

The search is performed using 2.04~$fb^{-1}$ integrated luminosity of data, collected at $\sqrt{s}$ = 7 TeV~\cite{WHlj}. The Higgs is produced in association with a $W$ boson, thus the final state includes at least two electron-jets and a $W$ candidate. Events are triggered by single electron or single muon triggers, using the lepton from $W \rightarrow l\nu$. The $W$ candidate is required to have an electron with $p_{T} >$ 25~GeV or a muon with $p_{T} >$~20 GeV, and $E^{miss}_{T} \geq$ 25~GeV.

The two required electron-jets are selected from candidate jets, and separated from SM jets based on their jet electromagnetic fraction $f^{EM}$, jet charged particle fraction $f^{CH}$, and the fraction of high threshold TRT hits $f^{HT}$ in their associated tracks. Prompt electron-jets are expected to deposit a high proportion of their energy in the electromagnetic calorimeter, to have an anomalously high charged particle fraction (since they should be composed entirely of charged electrons and positrons), and to have a large average number of tracks associated to the jet. Candidate electron-jets have $p_{T} \geq$ 30 GeV, $f^{EM} \geq$ 0.99, $f^{CH} \geq$ 0.66, and at least two tracks associated to the jet, where each track must have $f^{HT} \geq$ 0.08.

\begin{figure*}[htbp]
\includegraphics[width=0.39\textwidth]{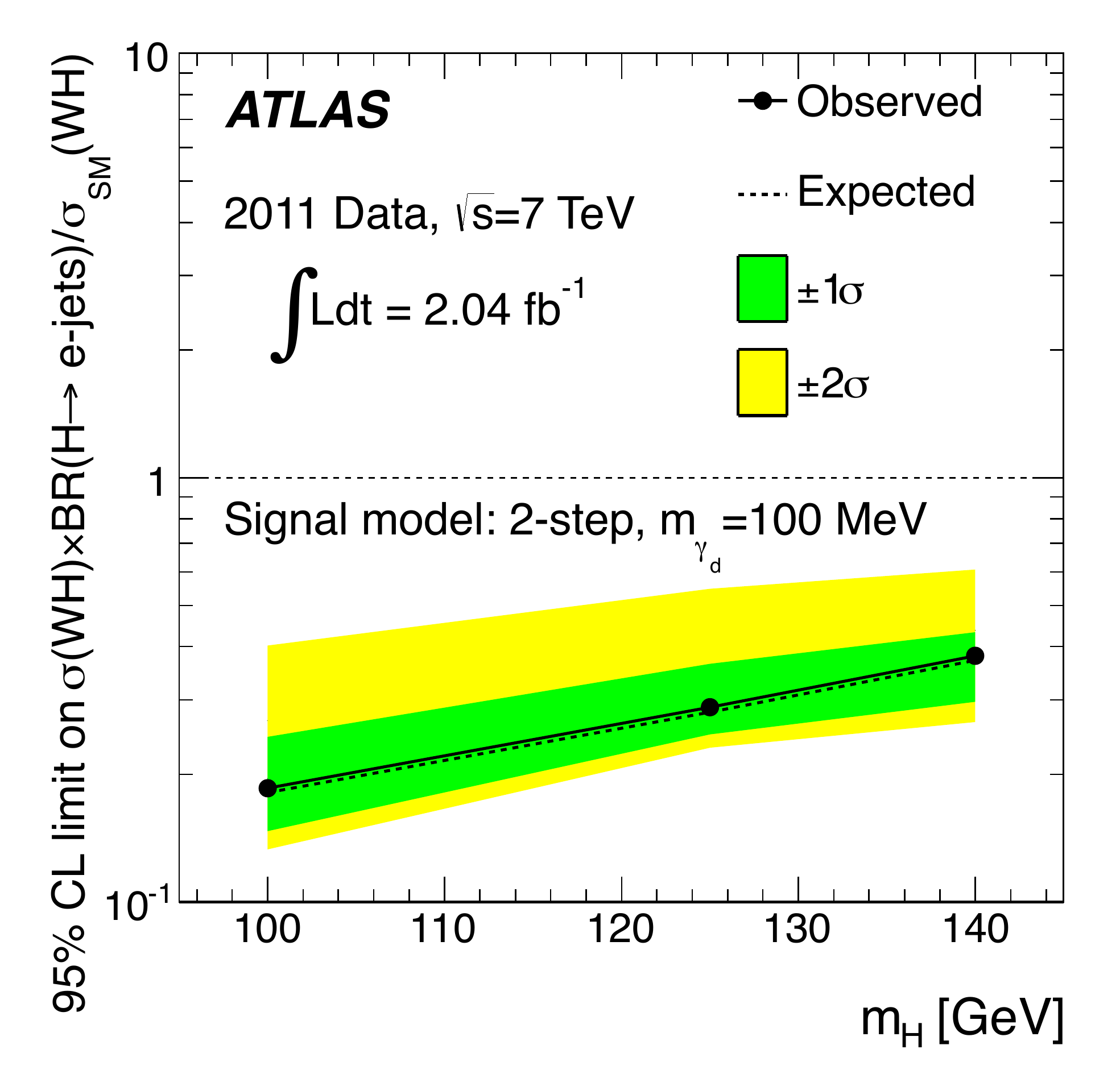}
\includegraphics[width=0.39\textwidth]{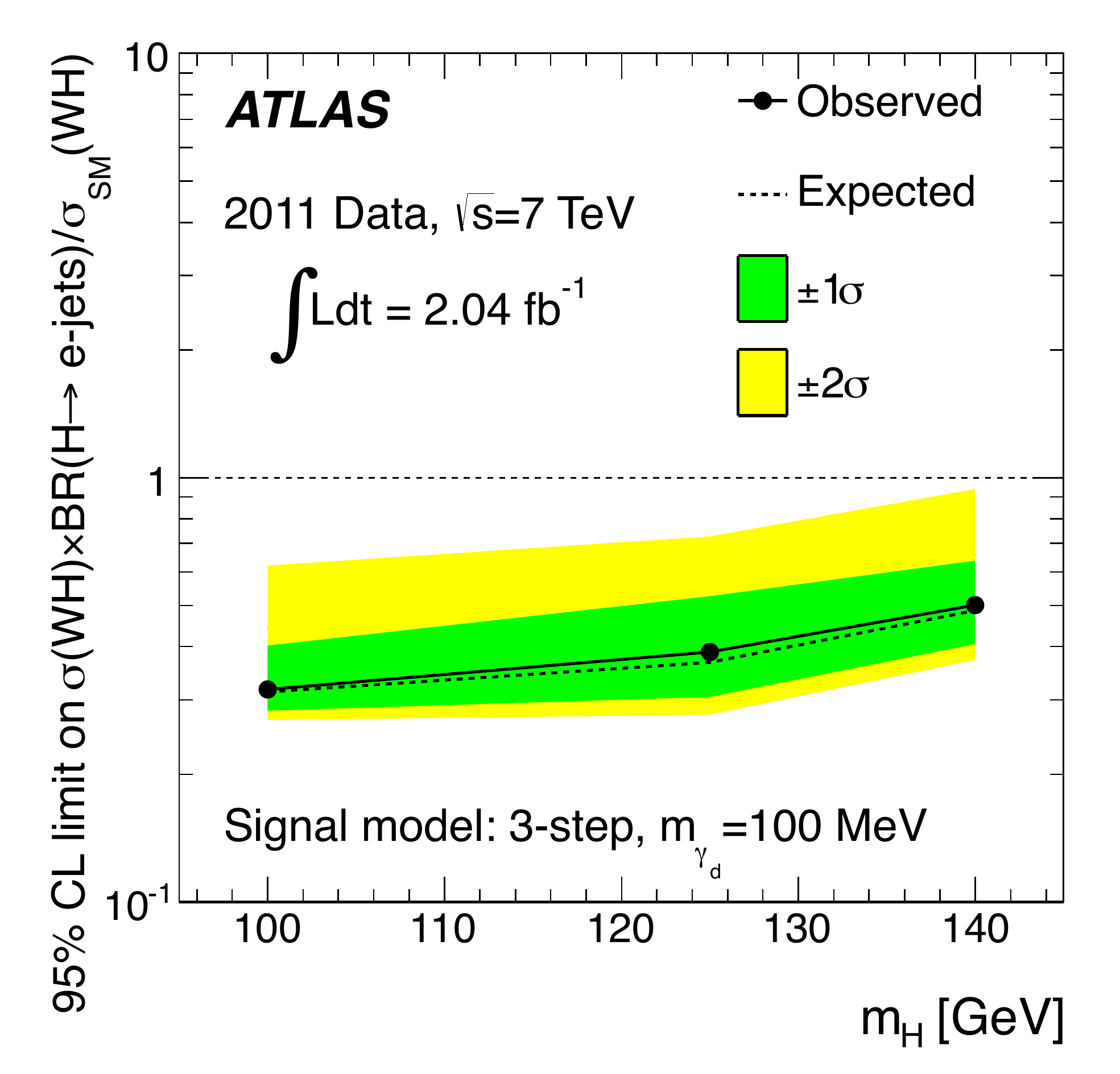}
\caption{Signal strength as a function of Higgs mass for a Higgs produced in association with a $W$ decaying to electron-jets. The left panel shows the limit for the process $H \rightarrow h_{d,1} \rightarrow h_{d,2} \rightarrow \gamma_{d} \rightarrow e-jets$, the right panel is for the process $H \rightarrow h_{d,1} \rightarrow h_{d,2} \rightarrow \gamma_{d} \rightarrow e-jets$~\cite{WHlj}.}
\label{WZljets}
\end{figure*}

Backgrounds include $W$+jets, $Z$+jets, $t\bar{t}$, diboson, and multijet production. The background contribution to the signal region is estimated using the matrix method, defining control regions by reversing or relaxing selection criteria and extrapolating the observed background contribution into the signal region. Less than one event is expected from background in the combined $W\rightarrow \mu\nu$ and $W \rightarrow e\nu$ regions, and zero events are observed in $W \rightarrow e\nu$, one in $W \rightarrow \mu\nu$. As the observed number of events is consistent with background expectations, limits are set on the signal strength, defined as $\sigma(WH)\times BR(H\rightarrow e-jets)/\sigma_{SM}(WH)$. The limits are shown in Fig.~\ref{WZljets} for two processes, $H \rightarrow h_{d,2} \rightarrow \gamma_{d} \rightarrow e-jets$ (the two-step process) and $H \rightarrow h_{d,1} \rightarrow h_{d,2} \rightarrow \gamma_{d} \rightarrow e-jets$ (the three-step process). Although the limits shown are for $m_{\gamma_{d}}$ = 100~MeV, the results for $m_{\gamma_{d}}$ = 200~MeV are compatible within uncertainties. The limits are translated into a limit on the Higgs branching ratio to electron-jets, excluded between 24 and 45\% for a 125~GeV Higgs at the 95\% CL.

\subsection{\label{sec:level2}Prompt lepton-jets from a squark decay in the Hidden Valley scenario}
A similar but distinct search was performed for prompt lepton-jets from a squark decay. In the relevant model, SM and SUSY are assumed, as well as a hidden sector of that is only accessible via interaction with the lightest supersymmetric particle (LSP). 

A squark pair is produced in $pp$ collisions, and the squarks each decay to a final state that includes the LSP. Lepton-jets are produced when the LSP decay produces a dark photon, which consequently decays to a lepton pair. In this measurement, both electron-jets and muon-jets are considered. In this scenario, depending on the dark sector gauge coupling $\alpha_{d}$, the dark photons can radiate more dark photons, resulting in lepton-jets with more than two leptons.

This search was carried out on 4.5~$fb^{-1}$ of data for the muon analysis and 4.8~$fb^{-1}$ of data for the electron analysis, collected at $\sqrt{s}$ = 7 TeV~\cite{HVlj}. Due to the possibility of dark-sector radiation, multiple classes of lepton-jet are defined. For the muon case, events are selected with at least two jets containing at least two muons, or at least one jet containing at least four muons. For the electron case, the calorimeter cannot resolve individual electrons, thus two jets are required, with no stipulations on the number of electrons in the jet.

Events are triggered with either single muon or single electron triggers. Electron-jet candidates are constructed from clusters in the electromagnetic calorimeter. The main discriminating variables for rejecting the multijet and $\gamma$+jet backgrounds are the cluster energy concentration, the electron cluster lateral width, the fraction of high threshold TRT hits, the calorimeter isolation, and the fraction of energy in the electromagnetic calorimeter. Electron jets are expected to have a high energy concentration, narrow lateral width, and to be isolated, due to their narrowness. As mentioned in the previous analysis, they are expected to deposit most of their energy in the EM calorimeter and to contain tracks with a large fraction of high threshold TRT hits.

Muon-jet candidates are reconstructed by seeding a jet finding algorithm with a high $p_{T}$ muon and adding muons within a cone of $\Delta R$ = 0.1 of the seed muon. The muon-jets are required to be well-isolated, and within the jet, the two muons most similar in $p_{T}$ must have an invariant mass of less than 2~GeV. Muon-jets are susceptible to background from the decays $\phi/\omega/\rho \rightarrow \mu^{+}\mu^{-}$.

For both the electron-jet and the muon-jet channels, the background contribution in the signal region is estimated using the ABCD method. In the electron-jet case, requirements on the EM fraction and electron cluster energy concentration were reversed to define the control regions. For the muon-jet case, selection on the lepton-jet isolation and $p_{T}$ of the third or fourth muon were reversed. No excess over the background prediction is observed in the signal region for the electron-jet channel. Slight excesses were observed for the single and double muon-jet channels, but with $p$ values of 0.06 and 0.04 for the probability that the background could fluctuate to the observed values. Thus, the observed yields are considered to be consistent with no signal, and limits are set on the  cross section times branching ratio. The limits are calculated for each value of $\alpha_{d}$ and $m_{\gamma_{D}}$, resulting in limits between 0.017 and 1.2 pb at the 95\% confidence level.

\section{\label{sec:conclusions} Conclusions}
Searches for signatures of exotic processes have been performed, probing a broad array of final state signatures. Even so, this represents only a subset of the beyond the Standard Model searches embarked upon at ATLAS. Although no sign of new physics has been observed, the recent measurements represent a substantial extension in experimental reach, and constrain the phase space available for theoretical models.




\end{document}